\documentstyle[aps,twocolumn]{revtex}
\begin{document}
\title{Charge distribution in C$_{60}$ crystal doped by electric field}
\author{
V.A. Kuprievich$^a$, O.L. Kapitanchuk$^{a,b}$, and O.V. Shramko$^a$}
\address{
$^a$Bogolyubov Institute for Theoretical Physics, 03143 Kiev, Ukraine\\
$^b$Kiev National Shevchenko University, Department of Physics,
03017 Kiev, Ukraine}
\maketitle

\begin{abstract}
The calculations of the charge distribution in C$_{60}$-based FET structure
are presented. A simple model is proposed to describe the distribution of
the injected electrons or holes between two-dimensional layers. The
calculations show that the relative charge distribution between the layers
turns to be independent on the total amount of injected charges. The charge
density is maximal on the surface layer and drops exponentially with the
depth increase. The relative amounts of injected charge involved in the top
layer are 73 and 64 per cent in the case of electron and hole injection,
respectively. Thus, the charge localization on the crystal surface turns to
be markedly different from near complete one that was obtained earlier
within tight- binding model for the charge concentration providing
superconductivity.
\end{abstract}


\section{Introduction}

Recently the superconductivity of organic substances due to injection of
charges by electric field in the field effect transistor (FET) was found
\cite{SKB00a,SKB00b,SKB01}. This new physical phenomenon is considered to be
unusual in several aspects. First, unusual is the nature of the
superconducting objects themselves: they are organic crystals, anthracene,
phenanthrene and fullerene C$_{60}$. Second, unusual is the
superconductivity of the charges injected by electric field: up-to-date the
superconductivity of organic crystals was related to the charge transfer,
particularly, due to chemical doping. Third, in the case of fullerene
C$_{60}$ the superconductivity is observed at extremely high temperature (52
K for pure C$_{60}$ \cite{SKB00a} and at even 117 K when doping by bromoform
\cite{SKB01}). Fourth, injecting holes into the fullerene leads to
significantly higher critical temperature than at injecting electrons,
although at chemical doping the C$_{60}$ superconductivity is observed with
negative charge carriers only.

These findings give rise to the problem of physical mechanisms of the
phenomena, in the first turn, including the question on the distribution of
charge carriers in organic crystal. In the study \cite{SKB00a} the amount of
induced charges per C$_{60}$ molecule is estimated with the supposition that
the additional charges occupy the surface monolayer only.

An almost complete surface localization of induced charges near
superconducting concentration is predicted in \cite{WPR01}. This study makes
use of tight-binding model taking into account electron repulsion by mean
field.  However, in the case of fullerite, the Hubbard repulsion $U_0$ on a
C$_{60}$ molecule is estimated to be 1 eV or more \cite{Gun97} whereas the
magnitude of the interfullerene hopping $t$ is below 50 meV \cite{WPR01}.
Thus, the fullerite crystal, which characterized by the relation $U/t$ more
than 20, have to be classified as a system with very strong electron
correlation. That is why the results obtained with the use of the
one-electron tight-binding model should be taken with care.

The above reasons motivate the present study of charge-profile problem in
fullerite.  We consider the problem from an opposite point of view that
seems to be in better correspondence with the electronic characteristics of
this material.  We apply the approach earlier used by us for calculations of
charge transfer in the fullerides A$_3$C$_{60}$ crystals \cite{KSK01}. It is
hopping smallness in comparison with electrostatic energies that is used to
develop a simple scheme allowing us to calculate the charge distribution in
the C$_{60}$-based FET structure.

\section{Model and method}

As in Ref.\cite{WPR01}, we consider the fcc fullerite lattice with a (001)
plane parallel to the gate. We are interesting in the distribution of
injected charges between the (001) parallel layers of C$_{60}$ molecules
that forms square lattices with the side length $b$=10~A.

Let $\rho_n$ be total number of extra electrons per molecule in layer $n$, so
their charge is $-e\rho_n$ (the case of injected holes is represented by
negative $\rho_n$). We consider the system with the total numbers of layers
$N+1$, the surface layer is labeled by zero.

The basic assumption of our model is neglecting the overlap between C$_{60}$
molecules. Accordingly, the total energy of the crystal is the sum of the
energies of the separate layers of molecules that interact
electrostatically. The energy $E_n$ per molecule in a layer $n$ with the
potential $U_n$ is
\begin{equation}
E_n =\sum_{n=0}^{N} [E_0(\rho_n) - eU_n \rho_n ],
\end{equation}
where $E_0(\rho_n)$ is energy of free molecule with $\rho_n$ extra
electrons.

This expression is accurate provided $\rho_n$ is integer, so $E_0(\rho_n)$
can be treated as energies of C$_{60}$ ions with corresponding charges. For
noninteger $\rho_n$, it is naturally assumed that $E_0(\rho_n)$ can be
interpolated by the quadratic fit,
\begin{equation}
E_0(\rho_n)=E_0(0)+E_1\rho_n + \frac{1}{2} E_2\rho_n ^2.
\end{equation}
Then the minimum condition of total energy with respect to $\rho_n$ leads to
equation
\begin{equation}
E_1 + E_2\rho_n - eU_n = \theta,
\end{equation}
where $\theta$ is a Lagrange multiplier taking into account the additional
restriction of fixing total electron number
\begin{equation}
\sum_n \rho_n =\rho_{tot}.
\end{equation}
From the other side, the layer potentials are, in turn, related to the
charges by the Poisson equations. Because of negligible hopping, as it is
adopted in our model, the C$_{60}$ molecules interact only electrostatically
and can be represented by point charges due to their near spherical form.
According to Ref.\cite{WPR01}, the potential $U_n$ of layer n is determined
by the equation
\begin{equation}
-eU_n=\xi \sum_{m=0}^{N} \min(m,n)\rho_m -\eta \rho_n,
\end{equation}
where the coefficients $\xi$ and $\eta$ are expressed through C$_{60}$
dielectric constant $\varepsilon$ and the distance $b$ between neighboring
molecules
\begin{equation}
\xi=\frac{4\pi e^2}{\varepsilon b\sqrt{2} },
\ \ \ \eta=\frac{3.9 e^2}{\varepsilon b}.
\end{equation}
Excluding $U_n$ from (3) and (5) one obtains the system of linear
inhomogeneous equations that determine, together with the condition
(4), the charge densities $\rho_n$,
\begin{equation}
\lambda \rho_n+\sum_{m=0}^{n} m \rho_m +n\sum_{m=n+1}^{N} \rho_n = C,
\ \ n=0,1\ldots, N,
\end{equation}
where parameter $\lambda$ is defined by the equality
\begin{equation}
\lambda=(E_2 -\eta)/\xi
\end{equation}
and $C=(\theta -E_1)/\xi$ is a new constant. As it can be seen from Eq.(7)
at $n=0$, this constant is in close relationship to the charge density in
surface layer,  $C=\lambda\rho_0$.

To treat the system (7), let us write it in the matrix form
\begin{equation}
({\bf A}+\lambda{\bf I}) \vec{\rho} = C{\bf V},
\end{equation}
where {\bf I} is unit matrix, {\bf V} is the column vector all the
components of which are units, and {\bf A} is $(N+1)\times(N+1)$ matrix of
form
\begin{equation}
{\bf A} = \left( \begin{array}{ccccc}
0 & 0 & 0 & 0 & \cdot
\\ 0 & 1 & 1 & 1 & \cdot
\\ 0 & 1 & 2 & 2 & \cdot
\\ 0 & 1 & 2 & 3 & \cdot
\\ \cdot & \cdot & \cdot & \cdot & \cdot
\end{array}
\right).
\end{equation}
To find $\rho_n$, let us solve the eigenvalue problem of the matrix {\bf A},
\begin{equation}
{\bf X}^+{\bf A}{\bf X} =  \epsilon, \ \ {\bf X}^+{\bf X} = {\bf I},
\end{equation}
where {\bf X} is square matrix the column of which are eigenvectors of {\bf
A} and $\epsilon$ is diagonal matrix of corresponding eigevalues. Then one
can obtain from (9) and (11)
\begin{equation}
\vec{\rho} = C{\bf X}(\epsilon+\lambda {\bf I})^{-1}{\bf X}^+{\bf V} .
\end{equation}

Eq.(12) shows that the shape of charge profile defined by the relative
charge distribution, is independent on $\rho_{tot}$. Obtaining $\rho_n$ at
any $C$, they are normalized to $\rho_{tot}$, thus satisfying the additional
condition (4).

\section{Numerical results and discussion}

The quantity $1-\rho_n$/$\rho_{tot}$ determines the relative part of
injected charge below the surface layer of the crystal. Its dependence on
the parameter $E_2$ is depicted in Fig.\ref{FIG1}, as the result of
calculations with $N$=50. We consider only range $E_2>\eta$  where the
solutions have physical meaning and all the $\rho_n$ has the same sign as
the constant C in (12).  The Fig.\ref{FIG1} shows that the charge density at
$(E_2-\eta)\rightarrow+$0 is completely localized on the surface molecules.
As $E_2$ rises, the charges markedly expands into deeper layers. To obtain
numerical estimations in real C$_{60}$ crystal one should evaluate only
$E_2$ in the approximation (2).  Considering the latter as a fit to the
ground-state energies of C$_{60}$ molecule and its ions one obtains the
following estimations of $E_2$ for injected charges of both signs
\begin{eqnarray}
E_2&=&E(0)+E(2)-2E(1)
\nonumber
\\&=&A(C_{60})-A(C_{60}^-)=2.7 eV,\ ({\rm for\ electrons})
\nonumber
\\ E_2&=&E(0)+E(-2)-2E(-1)
\nonumber
\\&=& -I(C_{60})+I(C_{60}^+)=3.8 eV,\ ({\rm for\ holes})
\nonumber
\end{eqnarray}
where $A$ are electron affinities and $I$ are ionization potentials with
known experimental values (in eV) \cite{Steg92,HS93,HCR91}:
\begin{eqnarray}
\nonumber
A({\rm C}_{60})&=&2.7,\ A({\rm C}_{60}^-) \approx 0,\\
\nonumber
I({\rm C}_{60})&=&7.6,\ I({\rm C}_{60}^+)=19.0-7.6=11.4.
\end{eqnarray}
Using the estimation of parameter $\xi$ in (6)  $\xi$ = 2.9 eV \cite{WPR01},
we obtain $\eta$ = 3.9$\xi \sqrt{2}/4\pi$ =1.27 eV. With above estimations
of $E_2$ the degree of charge delocalization  $1-\rho_n$/$\rho_{tot}$ is
evaluated as 0.266 for electrons and 0.359 for holes, as it is illustrated
in Fig.\ref{FIG1}. In other words, about one quarter of injected electron
charge and one third of hole charges turns out to be beneath the surface
layer of C$_{60}$ crystal.

The calculated charge distributions between the layers for electrons and
holes in the $N=50$ crystal is shown in Fig.\ref{FIG2}. It is clear that the
injected charge densities fall with the distance from the surface strictly
exponentialy in both cases, with the electron density dropping more rapidly
than hole one.

To conclude, we summarize the main differences of our results from obtained
in Ref.\cite{WPR01}. (i) The charge profile in \cite{WPR01} substantially
depends on the charge concentration whereas our relative distribution is
independent of it. (ii) According to \cite{WPR01} 98 \% of the total charge
is confined to the surface layer for doping $\rho_{tot}$ higher than 0.3, in
opposite to our calculations showing this degree less than 75 \% at any
doping.  The both results are in qualitative agreement with respect to
domination of injected charges in the top layer. At the same time, our
calculations show the possible nonnegligible occupation of undersurface
layers. The question on their role in the conductivity properties of
fullerite in FET structures needs in a separate consideration in future.
However, just now our results suggest that the existing estimations of
charge concentration providing the highest $T_c$ in superconductive FET
structures may be subject to refinement.

\begin{figure}
\caption{Relative occupation of undersurface layers,  $1-\rho_n$/$\rho_{tot}$,
vs curvature $E_2$ of energy-charge fit (2). Dashed lines indicate the
results of calculations for injected electrons and holes in C$_{60}$-based
FET structure.}
\label{FIG1}
\end{figure}

\begin{figure}
\caption{Charge density distributions (in logarithmic scale)
calculated for injected electrons and holes in C$_{60}$-based FET
structure.}
\label{FIG2}
\end{figure}

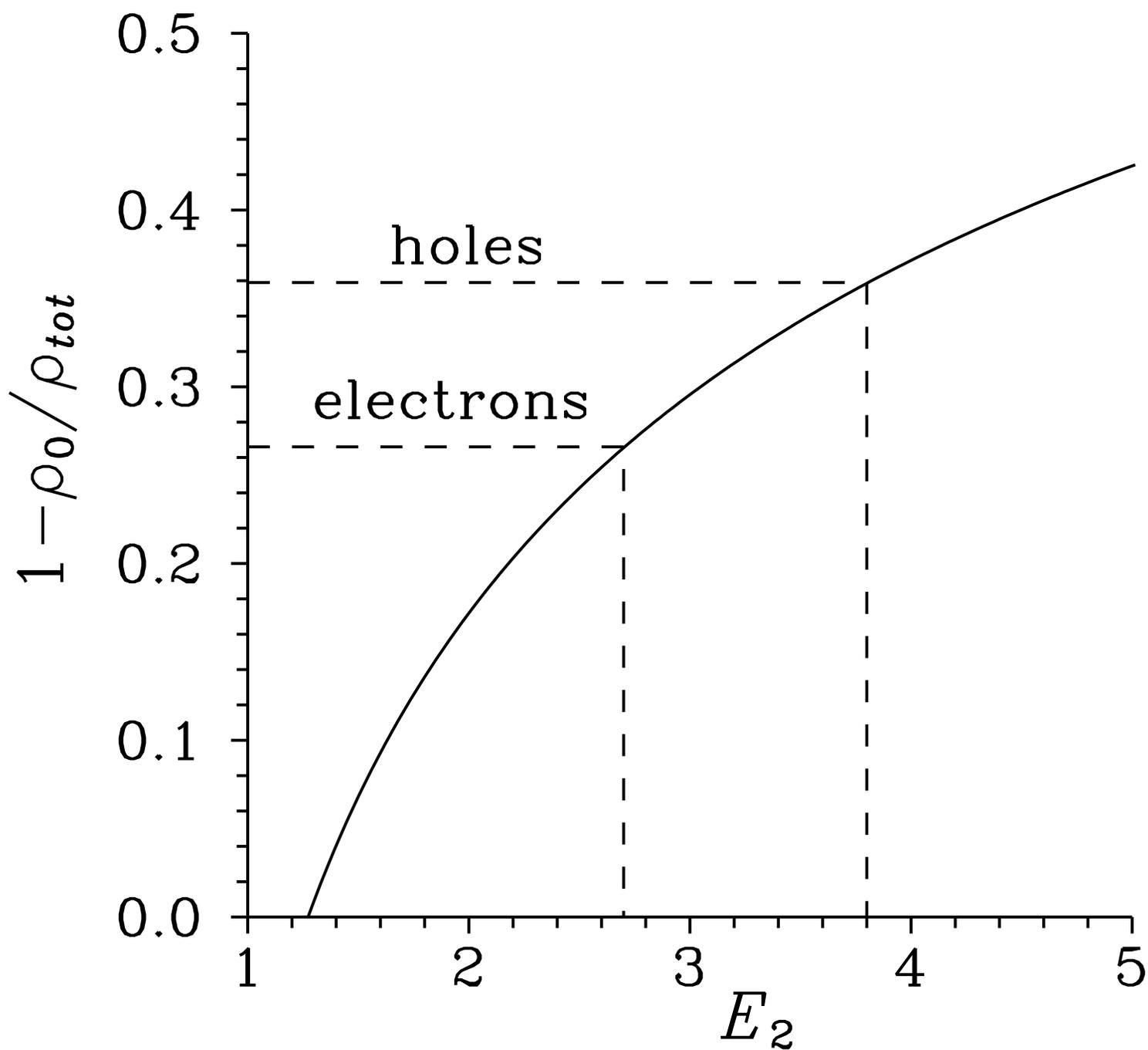
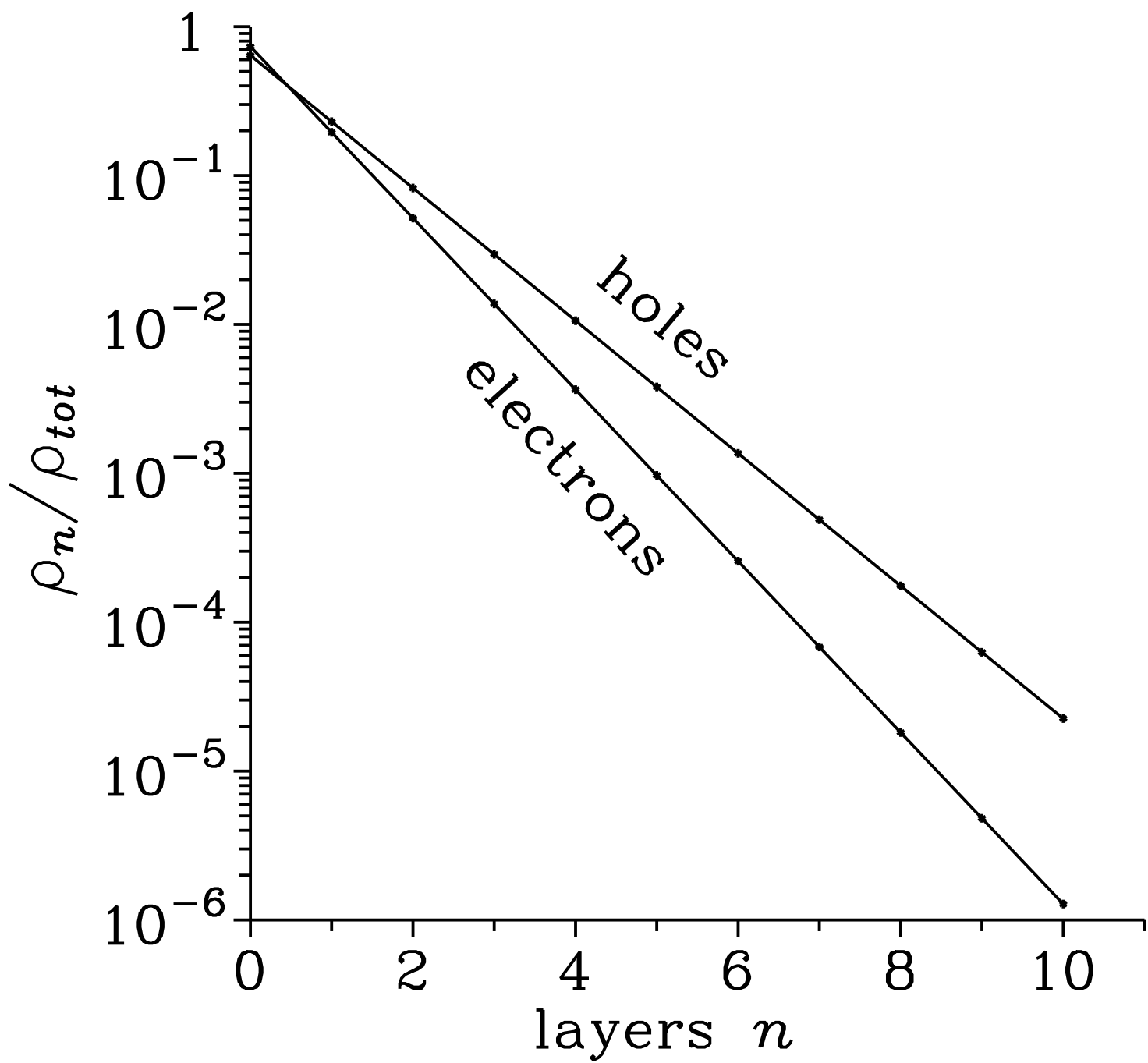
\end{document}